\documentclass[manuscript]{aastex}
\begin{document}

\def\msun{{$\,M_\mathrm{\odot}$}}
\def\rsun{{$\,R_\mathrm{\odot}$}}

\title{Infrared Spectroscopic Observations of the
Secondary Stars of Short Period Sub-Gap Cataclysmic Variables}

\author{Ryan T. Hamilton, Thomas E. Harrison}
\affil{New Mexico State University}
\affil{Department of Astronomy, PO Box 300001, MSC 4500, Las Cruces, NM 88003}
\email{rthamilt@nmsu.edu, tharriso@nmsu.edu}
\and
\author{Claus Tappert}
\affil{Universidad de Valparaiso}
\affil{Valparaiso, Chile}
\email{ctappert@astro.puc.cl}
\and
\author{Steve B. Howell}
\affil{NOAO}
\affil{950 N. Cherry Avenue, Tucson, AZ 85726}
\email{howell@noao.edu}

\begin{abstract} 

We present {\it K}-band spectroscopy of short period, ``sub-gap'' cataclysmic
variable (CV) systems obtained using ISAAC on the VLT. We show the infrared
spectra (IR) for nine systems below the 2-3 hour period gap: V2051 Oph, V436
Cen, EX Hya, VW Hyi, Z Cha, WX Hyi, V893 Sco, RZ Leo, and TY PsA.  We are able
to clearly detect the secondary star in all but WX Hyi, V893 Sco, and TY PsA.
We present the first direct detection of the secondary stars of V2051 Oph, V436
Cen, and determine new spectral classifications for EX Hya, VW Hyi, Z Cha, and
RZ Leo.  We find that the CO band strengths of all but Z Cha appear normal for
their spectral types, in contrast to their longer period cousins above the
period gap. This brings the total number of CVs and pre-CVs with moderate
resolution (R $\gtrsim$ 1500) IR spectroscopy to sixty-one systems: nineteen
pre-CVs, thirty-one non-magnetic systems, and eleven magnetic or partially
magnetic systems.  We discuss the trends seen in the IR abundance patterns thus
far, and highlight a potential link between anomalous abundances seen in the IR
with the \ion{C}{4}/\ion{N}{5} anomaly seen in the ultraviolet.  We present a
compilation of all systems with sufficient resolution IR observations to assess
the CO band strengths, and, by proxy, obtain an estimate on the C abundance on
the secondary star.  

\end{abstract}



\section{Introduction}\label{intro}

Cataclysmic variables (CVs) are short-period binaries in which a late-type,
Roche-lobe filling secondary star transfers matter through an accretion disk
onto a rotating, accretion heated primary white dwarf (WD). The standard
evolutionary paradigm \citep[hereafter HNR]{HNR} postulates that CVs evolve
from wide binaries of moderate orbital period and unequal masses. As the more
massive component evolves off of the main sequence into a red giant, the
secondary star finds itself orbiting within the atmosphere of the massive star.
During this common envelope phase, the orbit of the secondary star shrinks due
to interactions with the atmosphere of the more massive primary star. This
shortens the binary period, until the common envelope is ejected around orbital
periods of 1 day.  Angular momentum is then mainly lost through an efficient
magnetically constrained wind, \citep[``magnetic braking'', see][and references
therein]{Cameron2002}, shrinking the orbit so much that the Roche lobe of the
secondary star comes into contact with the stellar surface and mass transfer
begins, signaling the birth of a long period CV.  As the system evolves, the
secondary star continues to lose mass, but angular momentum losses keep the
secondary star in contact with its Roche lobe.  

During their lives as longer period CVs, orbital periods typically range from 3
to 10 hours, and mass transfer rates from $10^{-8}$ to $10^{-9}$ \msun
yr$^{-1}$. This rapid mass transfer drives the secondary star out of thermal
equilibrium, causing it to become bloated by $\sim$30\% compared to an isolated
star at the same mass \citep{Knigge}. The secondary continues to lose mass
until it reaches the fully convective boundary ($P_\mathrm{orb}\approx3$ hrs,
$M_\mathrm{2}\approx0.2-0.3$ \msun), where the magnetic breaking is believed to
be disrupted.  The secondary star is able to regain its thermal equilibrium,
and as a result shrinks and loses contact with its Roche lobe ceasing mass
transfer. With the disruption of magnetic breaking from the secondary star,
angular momentum in the system is lost via gravitational radiation alone,
causing the orbit to decrease at a much slower rate.  This continues until the
orbit has shrunk sufficiently for the secondary star to overflow its Roche lobe
and begin mass transfer once again ($P_\mathrm{orb}\approx2$ hrs).  With the
absence of mass transfer between orbital periods of 2-3 hours, the systems are
dormant, and consequently, difficult to identify.  This forms a gap in the
observed orbital period distribution of CVs between 2-3 hours.  Once mass
transfer resumes, the system emerges as a ``sub-gap'' CV.

It is assumed that this evolutionary sequence happens on a short enough
timescale that the secondary star does not undergo any significant nuclear
evolution, and therefore should retain normal abundance patterns consistent
with a main sequence dwarf.  Observational evidence, however, shows a growing
number of systems with apparent abundance anomalies seen in the UV and/or the
IR; in the UV, this is seen as unusual \ion{N}{5}/\ion{C}{4} ratios
\citep{Boris}, and in IR, this is inferred through the presence of anomalously
weak or absent CO features.  

\citet{TomLongPeriod, TomPolars, TomShorts} show that for thirteen out of the
nineteen systems observed, the CO features of non-magnetic CVs above the period
gap are much weaker than they should be for their spectral types.  Since the
water vapor features in the coolest of these stars appears normal, this result
points towards a deficit of carbon. In addition, $^{\rm 13}$C appeared to be
enhanced for several systems, an indication of CNO processed material in the
atmosphere of the secondary star \citep{TomLongPeriod}.  In contrast to these
non-magnetic CVs above the gap, the majority (8/11 systems) of ``polars'', CVs
with highly magnetic WD primaries, have secondary stars that appear to be
completely normal \citep{TomPolars, TomPolarsSpitzer}.  Pre-CV systems appear
uniformly normal, with only one of nineteen systems showing any weakened CO
features \citep{clausprecv, SteveRZLeo}.  While the total CV sample is somewhat
small and limited to the brightest objects observable with ground-based near-IR
spectroscopic instrumentation, these trends are striking enough to warrant
further attention and examination of all CV subtypes.

Observations of short period, sub-gap CVs are extremely challenging given the
very low luminosities expected for the secondary stars that must compete
against that of accretion and the underlying hot white dwarf. IR spectroscopy
of CVs has been possible since the early 1990's, notably with the efforts of
\citet{Dhillon95} looking at systems well above the 2-3 hour period gap, where
absorption lines from the secondary were clearly seen.  \citet{Dhillon2000}
observed five systems below the period gap, and found that the secondary stars
were too faint to detect and estimated that they contribute only 10 to 30\% to
the observed infrared flux.  

\citet{Mennickent} and \citet{MennickentRZ} focused on lower resolution studies
of sub-gap systems, fitting K or M dwarf template spectra to low resolution
optical and  near-IR spectra.  This technique was employed by
\citet{pasjshortcvs} who used the Subaru Telescope to obtain $J$, $H$, and
$K$-band low resolution grism spectroscopy of five CVs below the gap. Spectral
component fitting allowed them to obtain rough spectral type estimates ranging
from M1 to L1, but the low resolution of these data (FWHM $\sim 60$\AA)
prevented examination of the CO features.

To attempt to detect the secondary stars in a sample of sub-gap CVs, we have
obtained moderate resolution $K$-band spectra of nine systems.  This increases
the sample of CVs with moderate resolution (R $\gtrsim$ 1500) near-IR
spectroscopy to sixty-one systems: nineteen pre-CVs, thirty-one non-magnetic
systems, and eleven magnetic or partially magnetic systems.  Prior to this
work, three quarters of the non-magnetic systems were {\it above} the period
gap and only four below: RZ Leo, WZ Sge, GW Lib, and EI Psc.
\citet{SteveRZLeo} presented the $K$-band spectrum of RZ Leo, which showed no
evidence of a C deficit manifested as weakened or absent CO features.
\citet{TomVYEI} obtained moderate resolution observations VY Aqr and EI Psc,
finding evidence for strong deficits of C in both systems.  \citet{SteveWZSge}
presented observations of WZ Sge, showing this object to have both CO and H$_2$
emission from its accretion disk, the only such detection so far.

In our nine observed systems, we clearly detect the presence of the secondary
star in six of them. In two of the remaining cases, we are able to provide
constraints on their spectral types, and in only one system we could not detect
the signature of the secondary star. In contrast to the longer period CVs, the
majority of these secondary stars have CO features that appear to be present at
near-normal strengths.  We present our observations in \S\ref{obs}, the object
spectra and spectral type determinations in \S\ref{results}, a discussion of
our results in \S\ref{disc}, and our conclusions in \S\ref{conc}. 

\section{Observations}\label{obs}

Infrared spectroscopy for the program objects was carried out at the European
Southern Observatory (ESO) at Cerro Paranal, Chile, with the Infrared
Spectrometer And Array Camera (ISAAC) \citep{isaac} on the 8.2 meter Very Large
Telescope (VLT) Antu in service mode between May and August of 2008.
Additional infrared data were obtained for both RZ Leo and EX Hya utilizing
NIRSPEC \citep{NIRSPEC} at the W. M. Keck Observatory on Mauna Kea.  An
observation log describing the observations is presented in Table \ref{obslog},
listing the observation dates and times for each system, the exposure times,
the number of observations at each nodded position, and the percentage of the
orbital phase covered by the observations.

\subsection{VLT Antu}\label{vlt}

The ISAAC data were obtained in the standard infrared ABBA nodding pattern,
moving the object between two positions on the spectrograph slit to aid in the
removal of sky background.  These data were taken using the medium resolution
grating and a 0.6\arcsec\ slit, giving a resolving power of $R \sim 4400$ and a
dispersion of $1.20$ \AA\ pixel$^{-1}$ across the 1024 pixel square ISAAC
science CCD.  We used two wavelength centers ($2.25$ and $2.35\ \mu m$) giving
sufficient overlap to construct a composite spectrum covering $2.18$ to $2.40\
\mu m$. The observing conditions varied during the observation period but, in
general, these data were obtained in fair conditions with seeing generally less
than 1.25\arcsec\ as reported by the Differential Image Motion Monitor (DIMM)
at the VLT.

\begin{figure}
\epsscale{0.60}
\plotone{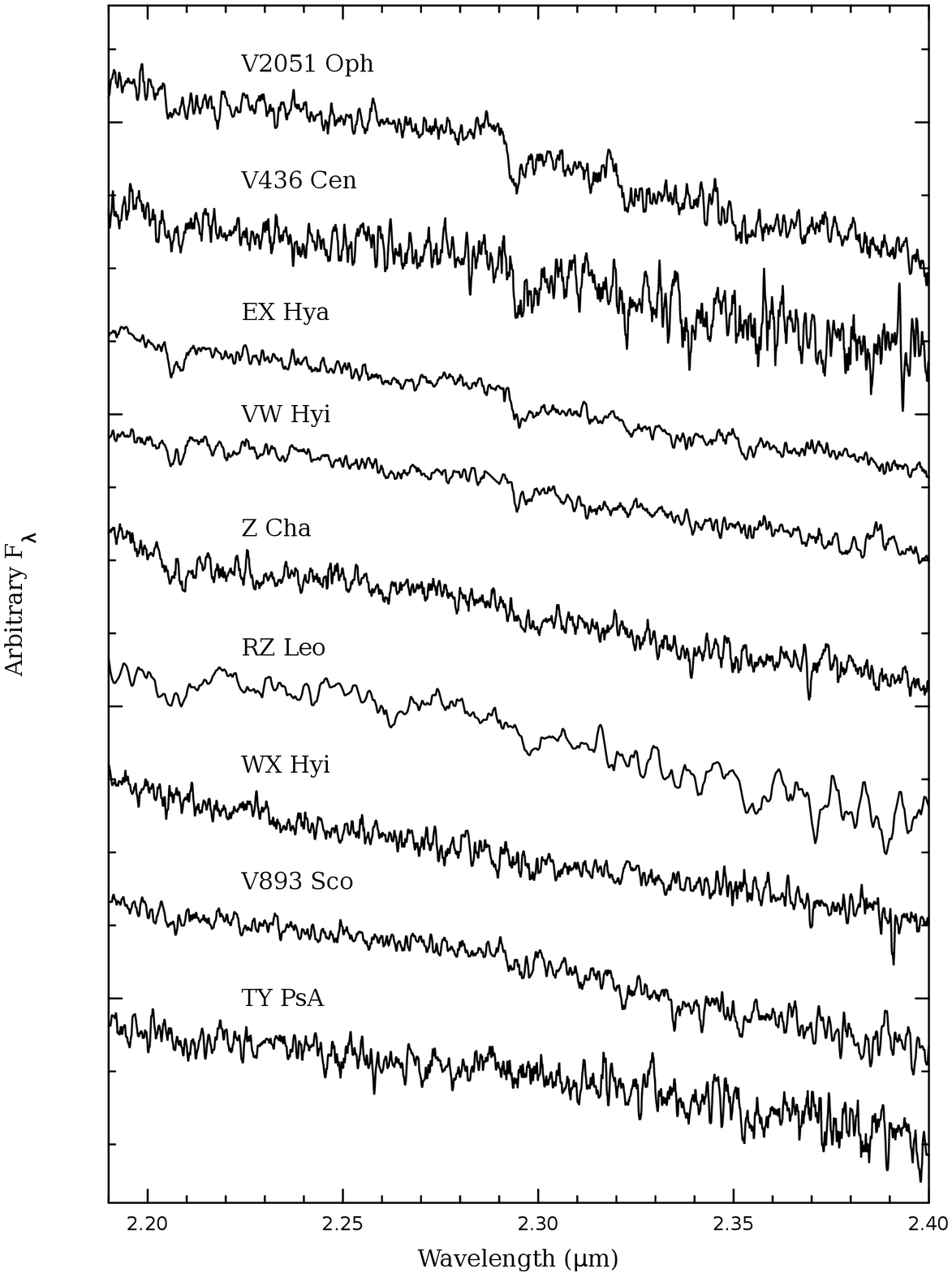}

\caption{The VLT spectra of our program objects, except that for RZ Leo which
was obtained using NIRSPEC on Keck II. }

\label{fig:figone}
\end{figure}

Several problems appeared during the observation run, however, that affected
the quality of these data.  Several large dust particles appeared on the
detector, degrading the cosmetics and compromising the spectral extraction.
This specifically impacted the data for V436 Cen and V893 Sco.  In addition,
significant drift in the central wavelength positions were observed,
particularly if the center position was changed before, or after, a telluric
standard was observed. Using night sky lines as a wavelength calibration source
mitigated this effect, but in general telluric correction and wavelength
calibration were more challenging than expected. The final spectra for our
program objects are shown in Fig. \ref{fig:figone}.

\subsection{Keck Observatory}\label{keck}

Both EX Hya and RZ Leo were observed using NIRSPEC at the W. M. Keck
Observatory, using a 0.380\arcsec\ slit and a low resolution grating covering
approximately 2.04 to 2.46 $\mu m$ with a dispersion of 4.27 \AA\ pixel$^{-1}$.
These data were obtained in the standard infrared ABBA nodding pattern, and
reduced using the IDL routine REDSPEC\footnotemark[1].  The data reduction
process followed the description given in \citet{TomPolars}.  Both objects had
telluric corrections applied in REDSPEC using observations of featureless A0V
stars close to the program objects to remove atmospheric absorption lines, and
to avoid any differences in telluric absorption dependent on airmass. We used
arc lamps to provide a wavelength calibration. The spectrum for RZ Leo has been
presented in \citet{SteveRZLeo}, but we include it here to compare it to our
results for the other sub-gap systems in Fig. \ref{fig:figone}.

\footnotetext[1]{Details about the REDSPEC IDL package can be found
on-line at the instrument website at
http://www2.keck.hawaii.edu/inst/nirspec/redspec.html.}

\subsection{Reductions}\label{red}

Standard data reductions of the VLT data were carried out using ESO
GASGANO\footnotemark[2] interface and the Common Pipeline Library (CPL) using
the ISAAC reduction recipes version 5.7.0.  The pipeline obtains a wavelength
calibration by examining the night sky lines in each data frame, and corrects
for field curvature using arc lamp frames collected at the end of each night as
well as dividing by a flat field. It then detects the spectrum present in each
input frame, shifting and co-adding to produce a single output spectrum.
Modifications were made to this pipeline to output the individual corrected
frames, which could then be properly combined after the reduction process to
account for the significant radial velocity motions of the secondary star.
While we were able to obtain a well corrected spectrum for each system, the
resulting radial velocity curves were extremely noisy and are not presented
here.  The modifications to the CPL recipes were checked for accuracy by also
examining the data with standard IRAF\footnotemark[3] methods ({\it APALL},
{\it IDENTIFY}, etc.) and found to give indistinguishable results. Telluric
features were removed using nearby A0V standard stars observed close to the
same airmass as the program objects.  Telluric standard stars were in general
observed at the end of each night, within $\pm\ 0.2$ airmasses of the program
objects.  In most cases only one telluric standard was observed each night.
These followed the same reduction process as the data frames, but instead used
the single combined output from the reduction pipeline since radial velocity
smearing was not an issue.  The most appropriate standard was then divided into
the observation using the IRAF task {\it TELLURIC}, which allows scaling and
shifting the standard to best match the atmospheric absorption lines present in
the data.  Care was taken to not scale or shift the standard too much and
introduce spurious absorption or emission lines to our program object spectra.

\footnotetext[2]{http://www.eso.org/sci/data-processing/software/gasgano/}

\footnotetext[3]{IRAF is distributed by the National Optical Astronomy
Observatories, which are operated by the Association of Universities for
Research in Astronomy, Inc., under cooperative agreement with the National
Science Foundation}

\subsection{M Star Templates}\label{mstars}

In addition to the spectra of the CVs, we obtained spectra of three late-type
dwarf templates: LHS 427, LHS 2347, and LHS 3003. Each reference object was
compared against standards taken from the IRTF Spectral Library
\citep{IRTFSTDs} as a check on the spectral reductions, and the spectral types
we derive are consistent with those reported in the SIMBAD database. The
observing log of the M stars observed as part of our program can be found in
Table \ref{obslog}. When necessary, these three M star templates were
supplemented with additional, lower resolution ($R \sim 2,000$) spectra from
the IRTF Spectral Library for comparison purposes.

\section{Results}\label{results}

In the following, we attempt to assign spectral types to the secondary stars of
our program objects if they are clearly visible. The \ion{Ca}{1} triplet at
$2.263\ \mu m$ and the \ion{Na}{1} doublet at $2.207\ \mu m$ are good
indicators of effective temperature \citep[and references therein]{Ivanov2004},
thus we use both as an indication of spectral type. As the effective
temperature decreases from M0, the \ion{Na}{1} doublet gets stronger and the
\ion{Ca}{1} triplet gets weaker.  The accretion disks in these systems are
bright, and provide a flat power-law continuum that veils the absorption lines
from the secondary star making a direct visual comparison to template spectral
types as described in \S\ref{mstars} more difficult.  To overcome this, prior
work \citep{MennickentRZ, Mennickent, pasjshortcvs} has focused on component
fitting a range of stellar template spectra with a variable power law component
to represent contributions from the disk to the {\it J},{\it H}, and/or {\it K}
band observations.

As a prelude to more in-depth synthetic spectral modeling currently underway,
we take a very simplistic, yet effective, approach to get first estimates of
the spectral types of the secondary stars visible in these data.  The visible
features in each object spectrum were enhanced using the IRAF task {\it SARITH}
to raise the spectrum to some constant power, and then fit from $2.18$ to
$2.28$ $\mu m$ with a linear function using {\it CONTINUUM} to remove the
overall continuum slope.  We do not fit the continuum over the entire spectral
range because of the presence of broad CO bands and changes in the continuum
shape due to the onset of strong water vapor absorption past $\lambda$
$\gtrsim$ 2.28 $\mu m$, both noted by \citet{clausprecv}. The very broad H$_2$O
features depress the continuum, especially in later spectral types, making it
difficult to ascertain its true level in the continuum fitting process.  The
region between the \ion{Na}{1} doublet and the \ion{Ca}{1} triplet, our primary
spectral type indicators, is relatively free of other contaminating sources, so
we are confident that the lines are undistorted by this process.  These object
spectra are then compared by eye to the IRTF spectral templates described in
\S\ref{mstars}.  The high accretion disk contamination and low S/N of some of
these data lead us to conclude that this approach was the best for determining
the spectral types of these secondary stars.  

\subsection{V2051 Oph}\label{v2051}

V2051 Oph ($P_\mathrm{orb}=1.50$ hr) is an SU UMa type CV, exhibiting
super-outbursts approximately every 227 days. It is also an eclipsing system,
and was observed by \citet{BAP98} using both ground based photometry as well as
$HST$ FOS spectroscopy.  They derived a secondary mass of
$M_\mathrm{2}=0.15\pm0.03$ \msun, and a radius of $R_\mathrm{2}=0.16\pm0.01$
\rsun, but make no estimate of the secondary spectral type. They do speculate,
however, that the system must be relatively young since the secondary does not
seem to be out of thermal equilibrium.  The empirical CV donor sequence by
\citet{Knigge} suggests, assuming a solar composition, that the secondary star
of V2051 Oph should have a spectral type near M7.  The 2MASS Point Source
Catalog \citep{2MASS} lists V2051 Oph as having $K_{\rm 2MASS}=13.53$.  

\begin{figure}
\epsscale{0.60}
\plotone{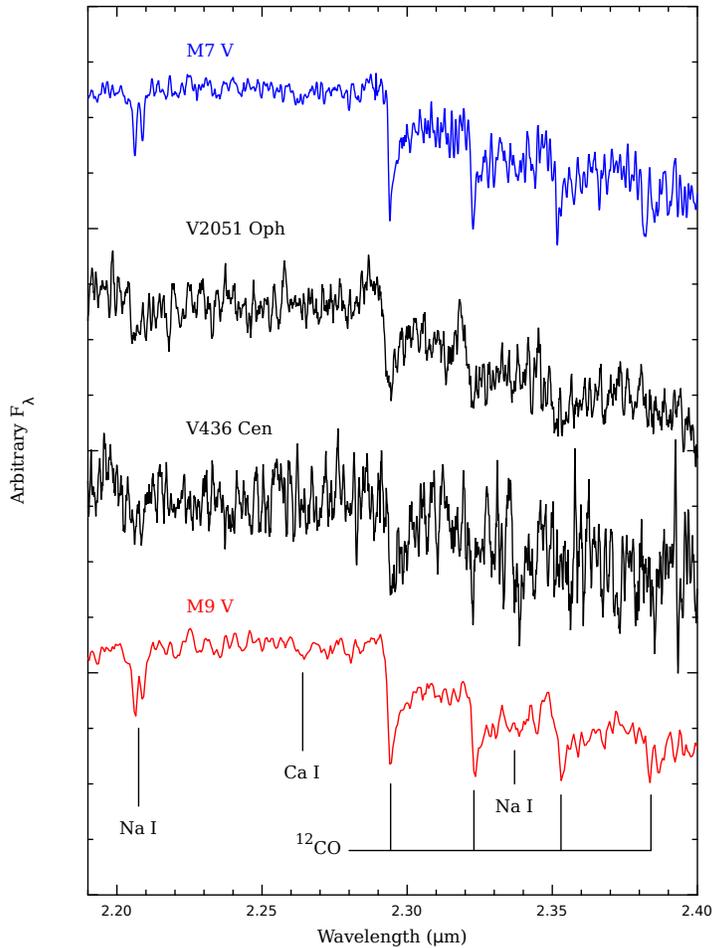}

\caption{The spectra of V2051 Oph and V436 Cen compared to the spectra of two
late-type templates an M7V (LHS 3003), and an M9V (LHS 2065, from the IRTF
Spectral Library). The strongest absorption features, as used for spectral type
determination, are indicated. The object spectra presented here have been
vertically stretched for presentation purposes to more clearly demonstrate
their similarity to the spectral type classification derived from the continuum
subtracted data as described in the text. The lack of Ca I absorption, and the
fact that the first overtone absorption of CO is stronger than the Na I
doublet, indicate a very late spectral type. We estimate spectral types of M7
for V2051 and M8 for V436 Cen.}

\label{fig:figtwo}
\end{figure}

The ISAAC spectrum of V2051 Oph is presented in Fig. \ref{fig:figtwo}. This
spectrum shows that the first overtone CO feature is stronger than the Na I
doublet, suggesting a very late spectral type. The \ion{Ca}{1} triplet is also
not clearly detected, as would be expected if the secondary had a late-M
spectral type. From comparison of its spectrum to the templates we derive a
spectral type of M7$\pm$1 for V2051 Oph. The secondary of V2051 Oph has not
been previously detected, thus this $K$-band spectrum provides the first direct
constraint on its spectral type.

\subsection{V436 Cen}\label{v436}

V436 Cen ($P_\mathrm{orb}=1.50$ hr) is an SU UMa type CV, exhibiting
super-outbursts approximately every 630 days.  It has been suggested
\citep{Patterson2001} that V436 Cen could harbor a low mass secondary based on
empirical fits to a relationship between the mass ratio $q$ and the superhump
and orbital periods.  In quiescence, V436 Cen has $K_{\rm 2MASS}=13.53$.
\citet{RitterKolb} report a secondary mass of $M_\mathrm{2}=0.17$ \msun, but
list it as uncertain.  It is interesting to note how similar V2051 Oph and V436
Cen are in terms of orbital period, $M_\mathrm{2}$, and even $K_{\rm 2MASS}$,
which all would imply that the secondary star of V436 Cen should be very much
like that of V2051 Oph, an M7 dwarf donor. 

The spectrum of V436 Cen, shown in Fig. \ref{fig:figtwo}, appears to be
slightly later than that of V2051 Oph. Using the Na I, Ca I, and CO features we
derive a spectral type of M8$\pm$1. The data reduction process for this object
was somewhat hampered by the lack of a good telluric standard to correct its
spectra. As with V2051 Oph, this is the first direct detection of the secondary
in V436 Cen.

\subsection{EX Hya}\label{ex}

EX Hya is a bright ($K_{\rm 2MASS}=11.69$), well studied short period
($P_\mathrm{orb}=1.64$ hr) Intermediate Polar (IP), and is the only magnetic CV
in our sample.  \citet{BeuermannEX} used optical spectroscopy and a previously
derived value of K$_1$, finding the mass of the secondary to be
$M_\mathrm{2}=0.108\pm0.008$ \msun, with a radius of
$R_\mathrm{2}=0.1516\pm0.0034$ \rsun. They assigned a spectral type of
M$5.5\pm0.5$ on the basis of \ion{Na}{1} and TiO band strengths compared to M4V
and M6V templates. $JHK$ spectra of EX Hya during outburst were presented by
\citet{TomMorePolars}, who noted that both water vapor absorption (at 1.38 and
1.9 $\mu$m) and the Na I doublet (2.2 $\mu$m) were visible even though EX Hya
was two magnitudes brighter than it is at quiescence.

\begin{figure}
\epsscale{0.60}
\plotone{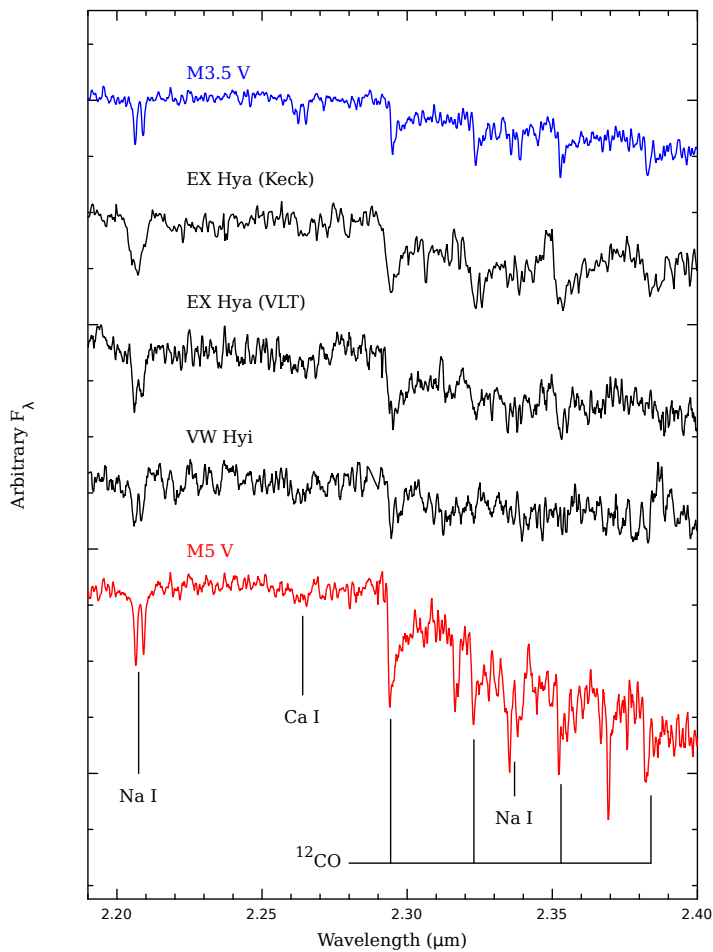}

\caption{The spectra of the intermediate polar EX Hya, and the SU UMa system VW
Hyi, compared to the spectra of two mid-M dwarfs (the M3.5V is LHS 427, and the
M5V template is LHS 2347). EX Hya was observed with both Keck (upper spectrum)
and with the VLT (lower spectrum). As in Fig. 2, the object spectra have been
stretched for presentation purposes. The relative strengths of \ion{Ca}{1}
triplets indicate spectral types near M5 for EX Hya, and M4 for VW Hyi.}

\label{fig:figthree}
\end{figure}

We have observed this system in quiescence with both ISAAC at the VLT, and with
NIRSPEC \citep{NIRSPEC} at the Keck Observatory.  Both of these spectra are
presented in Fig. \ref{fig:figthree}, where we compare them to the spectra of
two M dwarfs. In both datasets, the secondary star was prominent. The strength
of the Ca I triplet in this system indicates an earlier spectral type than
either of our first two objects, best matching an M5 dwarf. In contrast to its
longer period IP cousins, GK Per and AE Aqr \citep[see][]{TomPolarsSpitzer}, EX
Hya has CO absorption features that appear to be relatively normal. EX Hya has
a high precision $HST$ parallax \citep{EXparallax} that gives a distance of
64.5$\pm$1.2 pc. At this distance, if we assign a value of M$_K$ appropriate
for an M5 V dwarf, the secondary would supply 44\% of the observed $K$-band
flux.

\subsection{VW Hyi}\label{vw}

VW Hyi ($P_\mathrm{orb}=1.78$ hr) is an SU UMa type CV, with outbursts roughly
every 28 days and super-outbursts approximately every 183 days.
\citet{Mennickent} examined this system with ISAAC at low resolution, finding
evidence for an L$0\pm2$ dwarf based on fitting of the $J$-band with a
power-law disk component and a stellar template.  \citet{SmithVW} estimate a
primary mass of $M_\mathrm{1}=0.71^{+0.18}_{-0.26}$ \msun\ from the
gravitational redshift of the \ion{Mg}{2} $\lambda4481$ absorption line on the
white dwarf.  A mass ratio of $q=M_2/M_1=0.148\pm0.004$ from the
superhump-period excess \citep{Patterson1998} implies a secondary mass of
$M_\mathrm{2}=0.11\pm0.03$ \msun. \citet{Knigge} predicts an M5 dwarf donor at
this orbital period.  VW Hyi is one of the brightest of the (non-magnetic)
sub-gap CVs, having $K_{\rm 2MASS}=11.70$.

Our ISAAC spectrum of VW Hyi is shown in Fig. \ref{fig:figthree}. The strength
of the \ion{Na}{1} doublet relative to that of the \ion{Ca}{1} triplet point to
a mid-M spectral type, and we find good agreement with an M4 dwarf. At this
spectral type, the CO features are close to their expected strength.  Longward
of the first prominent CO bandhead around 2.29 $\mu m$ the spectrum becomes
very noisy, and the spike at the red end of the spectrum is due to low S/N, not
CO disk emission filling of the bandheads \citep[see][Fig.  11]{SteveRZLeo}.
Strangely, the spectral region between 2.21 and 2.26 $\mu m$ has a number of
absorption features, which we tentatively associate with Fe I, that are much
more consistent with a very late M-type dwarf (M9V).  While the S/N of this
spectrum is not particularly high, each of these features has the correct depth
and width (as some are doublets), matching the features in the M9V. Later
types, however, do not match the strengths of the other prominent spectral
features, and we find that the overall spectrum is clearly inconsistent with an
L0 dwarf.  

\subsection{Z Cha}\label{zcha}

Z Cha (P$_{orb} = 1.79$ hr) is an SU UMa type CV, with recurrence times of
$\sim 17$ days and $\sim 287$ days for normal and superoutbursts respectively
and $K_{\rm 2MASS}=13.31$. \citet{WadeZCha} observed Z Cha spectroscopically,
obtaining a radial velocity curve of the secondary star using the \ion{Na}{1}
doublet at $\lambda8183$ and $\lambda8194$. Combining that with a mass ratio of
the system derived from eclipses by \citet{WoodZCha}, they calculate system
parameters of $M_\mathrm{1}=0.84\pm0.09$ \msun\ and
$M_\mathrm{2}=0.125\pm0.014$ \msun.  \citet{WadeZCha} also find evidence for an
M5.5 secondary star, based on TiO band strengths. Since the orbital period for
Z Cha is so similar to VW Hyi, we would expect to find an M5 dwarf donor.

\begin{figure}
\epsscale{0.60}
\plotone{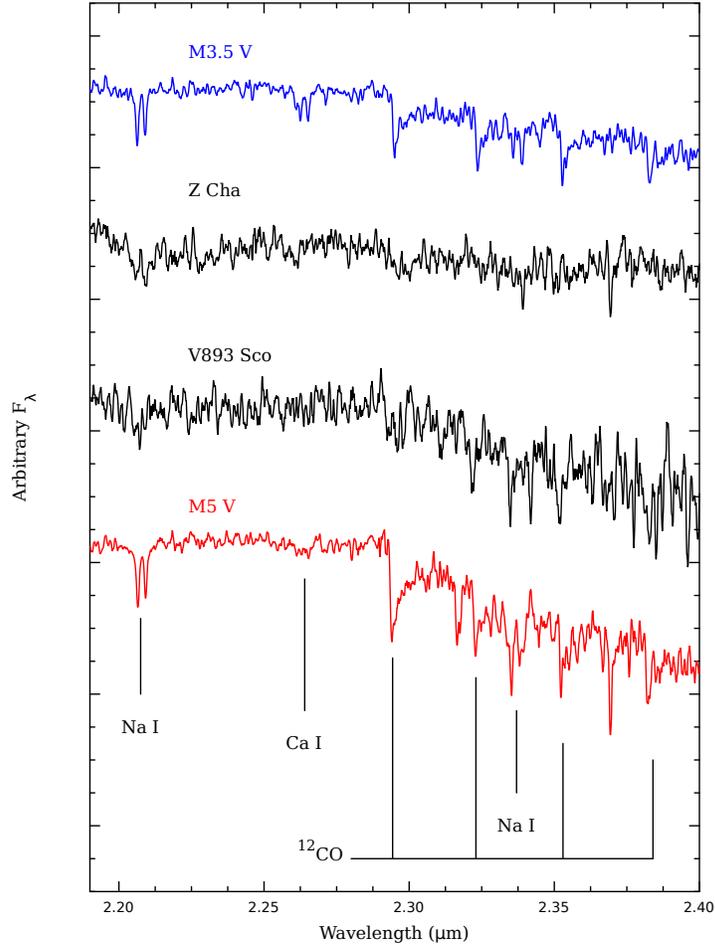}

\caption{The spectra of Z Cha and V893 Sco compared to two mid-M type
templates. In Z Cha, the \ion{Na}{1} doublet is clearly seen, as well as a hint
of the \ion{Ca}{1} triplet, suggesting a mid-M type secondary star. The CO
features, however, are quite weak, suggesting a C deficit. The spectrum for
V893 Sco is poorer, barely showing the Na I doublet, but it clearly has CO
absorption, indicating that it has a slightly later spectral type than Z Cha.}

\label{fig:figfour}
\end{figure}

The spectrum of Z Cha using ISAAC is shown in Fig. \ref{fig:figfour}. It must
be noted that the blue portion of the spectrum was observed some 4 months {\it
after} the redward side, as shown in Table \ref{obslog}.  The AAVSO archive has
magnitude estimates during both time periods, showing that the system was in
between outbursts at the epochs of the VLT observations, with similar visual
magnitudes. The spectrum is heavily contaminated by the accretion disk, making
proper identification of the spectral type of the secondary star highly
uncertain. If we accept the M5 classification, then the CO features are much
weaker than they should be.  This spectral type is consistent with the observed
strength of both the \ion{Ca}{1} triplet and the \ion{Na}{1} doublet.

\subsection{V893 Sco}\label{v893}

V893 Sco ($P_\mathrm{orb}=1.82$ hr) is an eclipsing SU UMa system with $K_{\rm
2MASS}= 12.68$.  \citet{MasonV893} obtained optical spectroscopy from which
they derived an H$\alpha$ radial velocity curve consistent with $M_\mathrm{1} =
0.89$ \msun\ and $q=M_2/M_1=0.19$, giving a secondary mass of $M_\mathrm{2} =
0.17$ \msun.  We again expect an M5 dwarf donor at this orbital period, as we
did for both VW Hyi and Z Cha.  \citet{Thorstensen}  was able to measure the
parallax of this system, putting the distance at $155^{+55}_{-34}$ pc. The Na I
doublet and the first overtone of CO are visible in our spectrum of V893 Sco,
but the quality of these data are low, and the Ca I triplet remains undetected.
This spectrum does show a water vapor break, and this is indicative of a later
type secondary than seen in Z Cha, and we suggest that the spectral type is $>$
M6. An M6V at the distance of V893 Sco would have $K=15.21$. Given that the
secondary star is visible in the $K$-band suggests that V893 Sco is closer than
given by the lower limit of the measured parallax.

\subsection{RZ Leo}\label{rz}

RZ Leo ($P_\mathrm{orb}=1.83$ hr) is a member of the small family of WZ
Sge-like CVs, sometimes called ``TOADs" \citep{SteveTOAD}, that exhibit
infrequent, but very large ($\ge 6$ mag) SU UMa-type outbursts.
\citet{Patterson2003} reports both an orbital and superhump period from
spectroscopy and long term photometry of 1.82492 and 1.888 hours, respectively.
\citet{Knigge} estimates, assuming a primary mass of $M_\mathrm{1} = 0.75$
\msun, an M5V secondary with a mass of $M_\mathrm{2}=0.114$ \msun\ and a radius
of $R_\mathrm{2}=0.171$ \rsun. This spectrum was recently presented by
\citet{SteveRZLeo}, and is included here for completeness. 

\begin{figure}
\epsscale{0.60}
\plotone{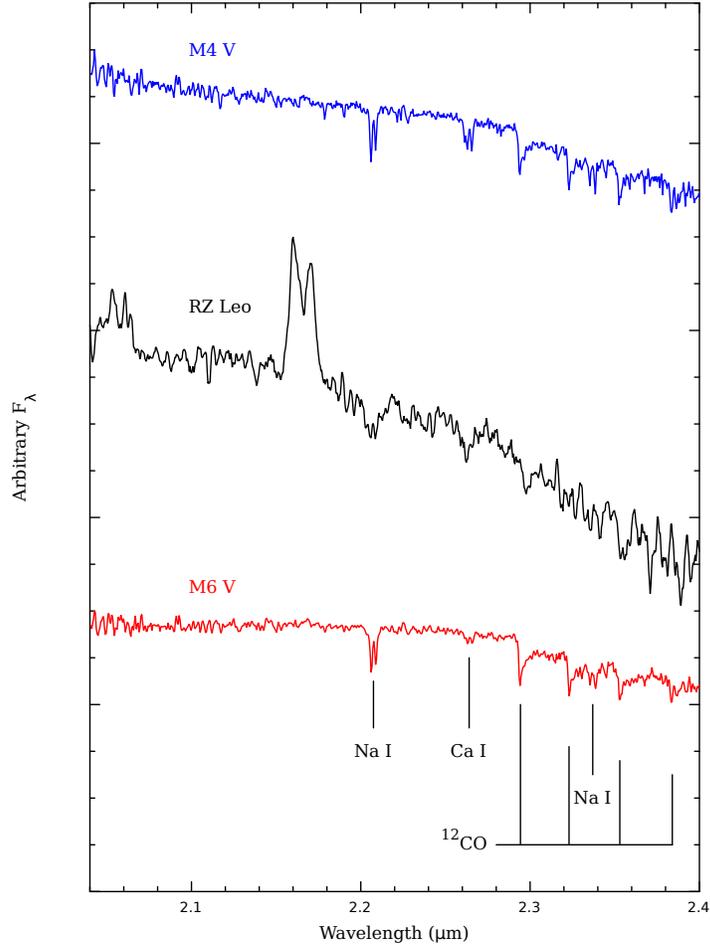}

\caption{The NIRSPEC spectrum of RZ Leo compared to two mid-M type dwarfs (both
from the IRTF Spectra library). The NIRSPEC data cover a larger range in
wavelength (at lower spectral resolution) than the ISAAC spectra. The two
strong emission lines are due to H I Br$\gamma$ (at 2.16 $\mu$m) and He I (at
2.06 $\mu$m). The (unstretched) spectrum of RZ Leo clearly shows a strong
decline due to water vapor.}

\label{fig:figfive}
\end{figure}

The {\it K}-band spectrum of RZ Leo from NIRSPEC at Keck as described in
\S\ref{keck} is shown in Figure \ref{fig:figfive}. RZ Leo is the faintest of
the objects in our survey (by two magnitudes), having $K_{\rm 2MASS}=15.39$.
Unlike the prototype TOAD WZ Sge, whose secondary has proved elusive, the donor
star in RZ Leo is easily seen. It appears to be an M4$\pm$1 with normal CO
features, based on the strengths of the \ion{Na}{1} doublet and the \ion{Ca}{1}
triplet. This spectral type is similar to the M5 assigned by
\citet{MennickentRZ} from SED fitting of lower resolution ISAAC {\it J}, {\it
H}, and {\it K}-band spectra. 

\subsection{WX Hyi and TY PsA}\label{nulls}

WX Hyi ($P_\mathrm{orb}=1.80$ hr) is an SU UMa type CV, and was first
spectroscopically examined by \citet{Schoembs}.  Using optical spectroscopy,
they found a primary mass of $M_\mathrm{1}=0.9\pm0.3$ \msun\ and mass ratio $q=
M_2/M_1=1.8$ based on a radial velocity study.  Ritter \& Kolb used these to
estimate a secondary mass of $M_\mathrm{2}=0.16\pm0.05$ \msun. We have been
unable to find any other information about the possible spectral type of the
secondary star. Even though it is relatively bright, $K_{\rm 2MASS}=12.96$, no
identifiable absorption features can be seen in its infrared spectrum (see Fig.
\ref{fig:figone}). In fact, in contrast to every other object, it shows a
slightly rising continuum at the red end of the $K$-band. 

TY PsA ($P_\mathrm{orb}=2.02$ hr) is a rarely studied SU UMa type CV.  Prior
spectroscopic examinations of this system \citep{TYPsA1,TYPsA2} found no radial
velocity variations, suspecting that this system has an extreme mass ratio and
therefore a sub-stellar companion.  \citet{Mennickent} examined this system in
the infrared with ISAAC previously, and found no features of the secondary
star.  At this slightly longer orbital period, \citet{Knigge} predicts an M4.5
dwarf donor.  Unlike WX Hyi, the continuum for TY PsA shows a slight decline at
the red end of its spectrum. Given the considerable contamination needed to
wash-out the other absorption features, this would suggest a late spectral type
($>$ M6) so as to have a large enough water vapor feature to affect the
continuum.  Given that it has an almost identical brightness to V2051 Oph
($K_{\rm 2MASS}=13.58$ vs. $13.53$), and we used the same exposure time as on
that object, the 2MASS survey must have caught this object somewhat brighter
than it is at true quiescence. Using our raw data shows that the count rate for
TY PsA was only half that of V2051 Oph, suggesting that at the time of
observation $K \approx 14.3$.

\section{Discussion}\label{disc}

\begin{figure}
\epsscale{1.0}
\plotone{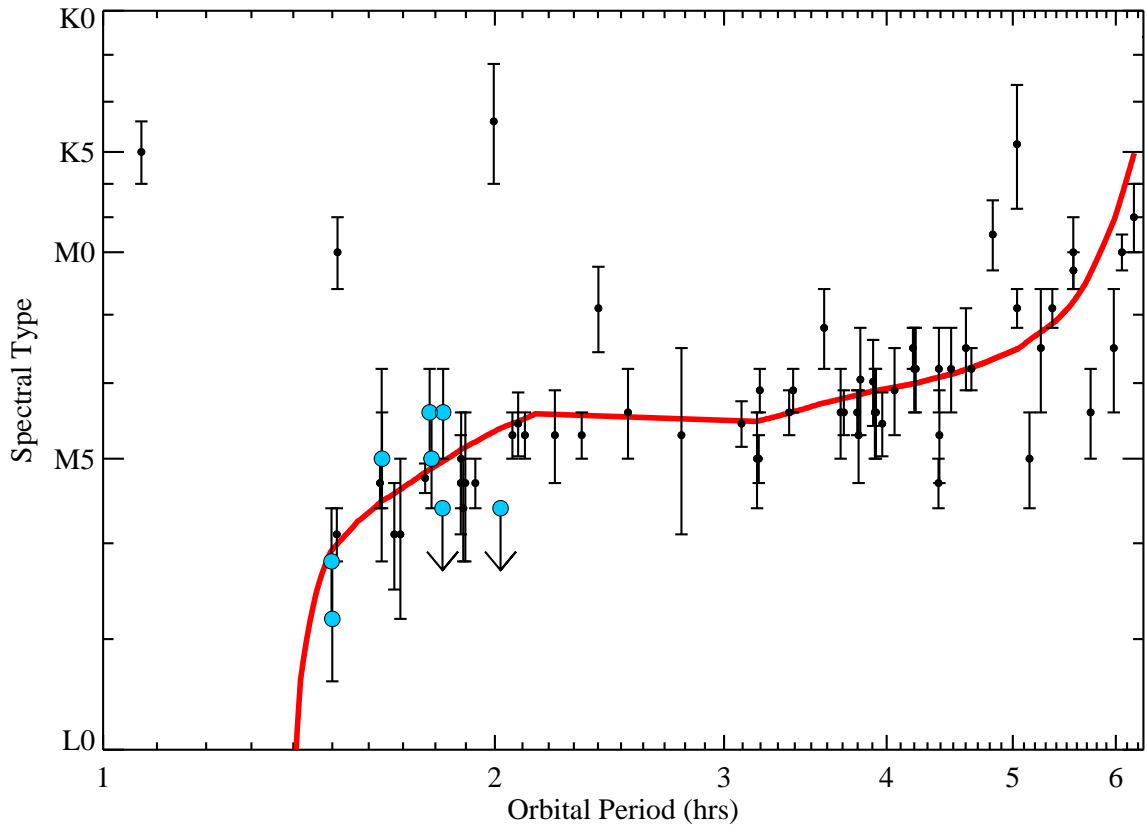}

\caption{The Spectral type vs. $P_\mathrm{orb}$ relationship from
\citet{Knigge} with the additional results from the analysis of the VLT ISAAC
spectroscopy plotted as larger circles (blue in the online version).}

\label{fig:figsix}
\end{figure}

We have conducted a moderate resolution spectroscopic survey of nine CVs below
the period gap. We clearly detect features from the secondary stars in six of
these objects, and place constraints on two systems based on suspected water
vapor declines seen in their continua.  All of the secondary stars appear to be
mid- to late-type M dwarfs and are consistent with the measured/estimated
masses available in the literature.  Our results are summarized in Table
\ref{ourresults}. In Figure \ref{fig:figsix} we have plotted our new spectral
type estimates on the empirical donor sequence \citep{Knigge}.  We have also
added the results from \citet{TomVYEI} for EI Psc and VY Aqr to this diagram
for completeness. Overall we find that our estimates fit well with Knigge's
empirical fit.

\begin{figure}
\epsscale{0.60}
\plotone{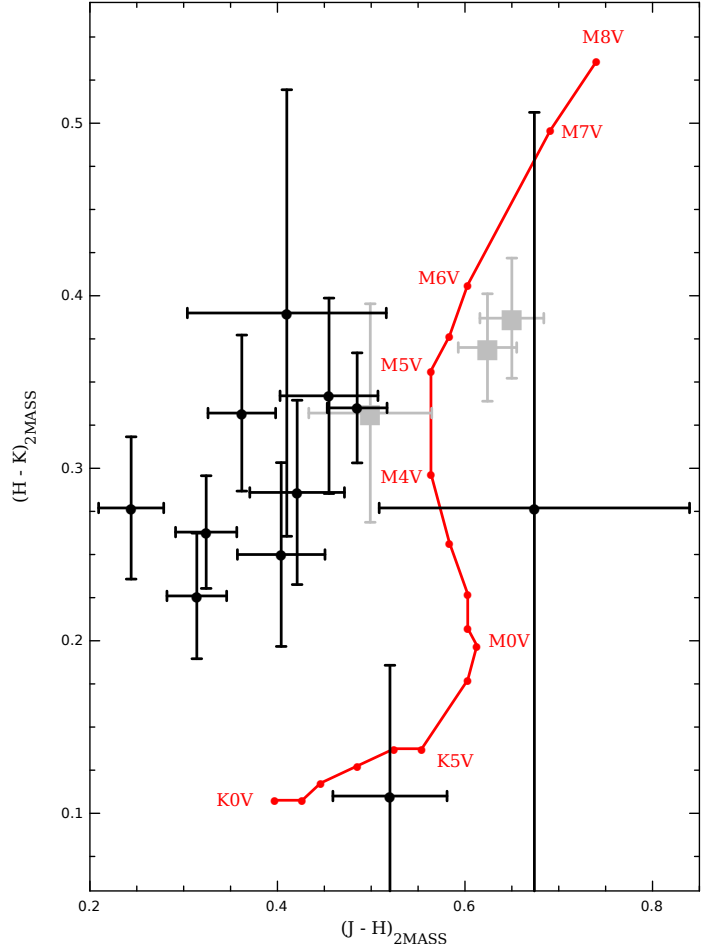}

\caption{The infrared color-color plot for the program objects generated using
the 2MASS data base. In addition, we have added two additional sub-gap CVs with
detected secondaries, EI Psc and VY Aqr, using data from \citet{TomVYEI}.  
Also shown is the main sequence color-color relationship (labeled line)
from K0V to M8V, and the reference M stars (see \S\ref{mstars}) observed in our 
sample (grey boxes).}

\label{fig:figseven}
\end{figure}

With the results of this diagram in hand, \citet{Knigge} discussed how to use
it to provide lower limits on the distance using single epoch $K$-band
measurements, with the warning that the distance is on average underestimated
by a factor of 1.75.  By combining our spectroscopically determined spectral
types with the observed infrared photometry, we find that this procedure can
only provide a weak constraint on the distances to sub-gap CVs.  In Figure
\ref{fig:figseven} we plot the 2MASS colors for our program objects, along with
the main sequence color-color relationship. Two objects fall near the main
sequence relationship: EI Psc, and RZ Leo. As discussed in \citet{TomVYEI} EI
Psc ($P_\mathrm{orb}=1.07$ hr) acts just like a main sequence K5 dwarf in the
infrared, and apparently suffers from very little accretion disk contamination.
The other object, RZ Leo, has such uncertain photometry ($\sim \pm 0.2$ mags,
see Table \ref{sysparams}) that its position in the diagram is very poorly
constrained and does not allow for any conclusions.  The objects with the
best-determined spectral types in our survey are V2051 Oph, V436 Cen, VW Hyi,
and EX Hya. The first three of these objects have very similar colors
(especially in $H-K$), but range from M8 (V436 Cen) to M4 (VW Hyi). This result
suggests that there is substantial disk contamination occurring in both V2051
Oph and V436 Cen --- yet these two objects have secondary stars that are just
as easily seen as those in VW Hyi and EX Hya.  It must also be noted that all
of these objects are capable of showing quiescent variability on the order of
$\Delta$V = $0.1 - 0.3$ mag.  
 
To underscore the problems associated with estimating parameters from IR colors
alone, we consider the two systems in our survey with measured parallaxes: EX
Hya, and V893 Sco. Using the Knigge relationship for EX Hya, one derives a
lower limit to its distance of $d=30\ pc$, while the parallax gives 62 $pc$.
The relationship would predict a distance of 53 $pc$ for V893 Sco versus the
parallax measurement of 155 $pc$. Both of these limits are less than one half
of what is measured, even accounting for the significant uncertainty in the
parallax for V893 Sco.

\begin{figure}
\epsscale{0.60}
\plotone{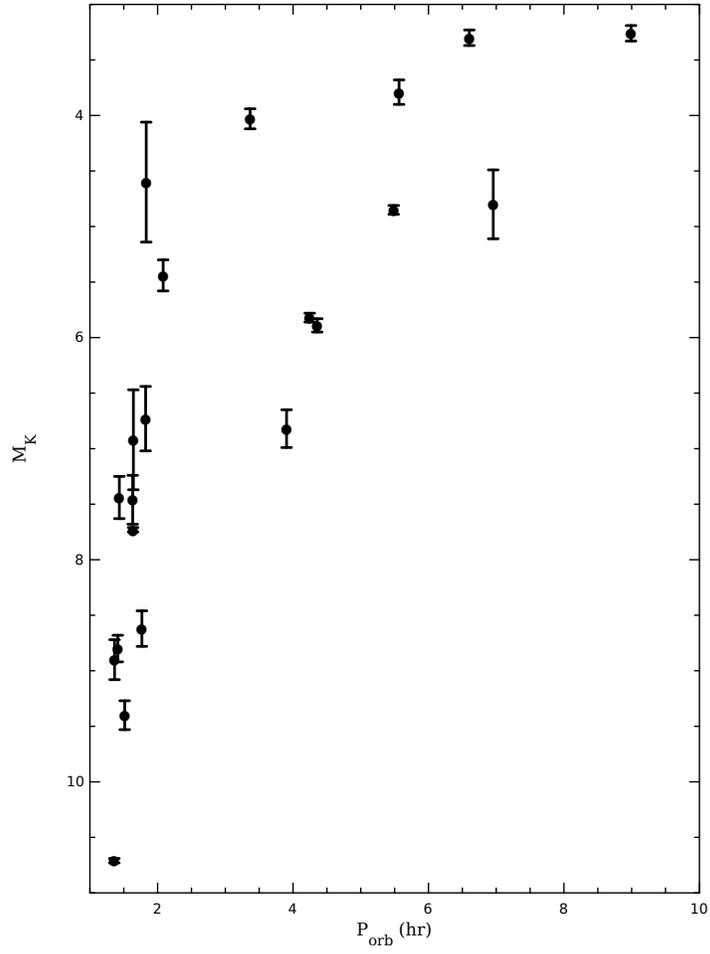}

\caption{The $M_\mathrm{K}$ vs. $P_\mathrm{orb}$ relationship for cataclysmic
variables with measured parallaxes (excluding ``polars'' and AM CVn systems).}

\label{fig:figeight}
\end{figure}

We show this further by assembling all of the published parallaxes for CVs (but
excluding polars, and AM CVn systems) and derive an $M_{\rm K} -
P_\mathrm{orb}$ relationship \citep[parallaxes from][]{TomParallax,
Thorstensen, Thorstensen08}, presented as Fig.  \ref{fig:figeight}. While there
is clearly a lower limit on M$_{\rm K}$ that depends on $P_\mathrm{orb}$, {\it
at any one orbital period the spread in M$_{\rm K}$ is two magnitudes or more}.
Deconvolving the infrared colors of CVs remains a difficult task, and infrared
photometry should be used with caution when estimating {\it any} intrinsic
parameter of these systems.

\subsection{Examination of CO Strength Across All CV Subtypes}

Table \ref{biglist} summarizes all the relevant observations at a high enough
resolution in the $K$-band to directly measure absorption lines from the
secondary star in all CV subtypes completed thus far.  Most of the data in
Table \ref{biglist} have been compiled from \citet{SteveRZLeo}, and we have
included those data here after checking on the original references for each
system.  While we are dealing with a small sample size, a few general trends
are clear: pre-CVs, sub-gap systems, and highly magnetic polars generally show
normal CO band strengths.  Eighteen of the nineteen pre-CV systems observed
show CO appropriate for their spectral type, with the exception of the triple
system HS1136 which showed no absorption features from the secondary star,
likely due to the presence of an extremely hot WD in the system. Among the
magnetic CV's, only three of the eleven show unusual CO strengths: GK Per, AE
Aqr, and V1309 Ori.  Both AE Aqr and GK Per are thought to harbor subgiant
secondary stars \citep{TomPolars, SteveRZLeo}, so this is not unexpected.
Among the non-magnetic, long period systems, there are thirteen out of nineteen
systems that either show absent or weakened CO features, and five systems that
appear more or less normal. Including the systems presented here, three out of
the twelve short period sub-gap systems clearly show weaker CO than expected
for their spectral type: Z Cha, VY Aqr, and EI Psc.  There is strong evidence
for a hot K-type secondary in EI Psc \citep{ThorEIPsc}, unexpected given its
ultra-short period ($P_\mathrm{orb}=1.07$ hr). It should have a very late type
companion, but the IR spectroscopy confirms the mid-K spectral type, and
reveals a secondary star with extraordinarily weak CO features. The results for
VY Aqr ($P_\mathrm{orb}=1.51$ hr) are not as clear, as the spectral type of the
secondary star derived from the VLT ($\sim$ M6) and Keck ($\sim$ M0) datasets
are quite different. No matter the classification, the CO features in the
secondary star of VY Aqr were extremely weak. For Z Cha, the case is not as
strong, though it appears that the CO strength is weaker than would be
expected.  Further observations are needed to verify the spectral type of the
donor star in Z Cha.

In all the cases where we see weak or non-existent CO absorption in the
atmosphere of the secondary star, we interpret this as a deficit in the C
abundance.  While we have IR spectra for a small percent of all CVs, UV
spectroscopy is uncovering evidence of C depleted material through the
detection of unusual N V/C IV line ratios. Studies in the UV
\citep[e.g.,][]{BBMouche, Szkody96, Mauche97} have long shown that the
\ion{N}{5}/\ion{C}{4} ratio in the UV spectra of a number of cataclysmic
variables was very high, suggesting a strong enhancement of nitrogen and/or a
deficit of carbon in the accretion disk or in the WD photosphere.  UV
observations of a large sample of polars by \citet{Betancor}, however showed
that they had normal \ion{N}{5}/\ion{C}{4} line ratios.  These two taken
together imply that matter transfered from the secondary star in non-magnetic
systems is the source of the UV abundance anomalies, and that these stars have
CNO processed material in their atmospheres.

Table \ref{biglist} highlights the five cases where we do have deficiencies in
C seen both in the UV and in the IR: EY Cyg, EI Psc, AE Aqr, U Gem, and V1309
Ori.  The most natural explanation of this result is that C depleted material
is being transfered from the secondary star to the white dwarf/accretion disk.
\citet{SteveRZLeo} provide a discussion of why the C deficient material must be
flowing from the secondary star onto the white dwarf/accretion disk and not in
the other direction.

The link between the abundances seen in the UV and the IR can be further
strengthened by looking at two cases where normal C abundances in the UV are
matched by normal C abundances in the IR: SS Aur and VW Hyi. SS Aur is a
typical dwarf nova above the period gap ($P_\mathrm{orb}=4.26$ hr), and appears
to have normal CO features in the spectrum presented by \citet{SteveRZLeo}.
Analysis of the UV spectra of SS Aur by \citet{GodonSSAUR} show that it has
normal C and N abundances.  Our spectrum of VW Hyi presented here show it to be
a normal M4 dwarf, and while previous UV spectroscopy suggested a sub-solar C
abundances \citep{SionVW}, more recent analysis (E. Sion, private
communication) suggests that the C abundance in VW Hyi is close to solar. To
put this potential UV-IR CNO connection to the test, it should be seen whether
an anomaly in one can be used to predict an anomaly in the other.  It is
therefore critical to obtain IR spectra of the three sub-gap systems that are
known to show unusual UV C/N ratios: BC UMa, SW UMa, and BW Scl \citep{Boris}.

It is a challenge to completely explain why pre-CV, short period, and magnetic
systems in general appear to have normal C abundances, while the long period
systems do not. One possible way to explain ``normal'' C abundances at short
periods was demonstrated by \citet{MarksSarna}. For their models where the
secondary has evolved off of the main sequence before contact, they show that
the surface C abundance in the secondary star declines throughout most of its
life as a CV (see their Fig. 16) until its mass reaches $\sim 0.3$ \msun.
After this point the surface C abundance returns to normal as material from
deeper layers within the star, unaffected by the CNO cycle that operated prior
to contact, are convected to the surface.  However, the standard model for CV
evolution does not give the secondary star sufficient time to evolve off the
main before starting mass transfer.  Still, if the angular momentum braking
mechanisms are less efficient than is currently believed, then the secondary
would have additional time to undergo nuclear evolution.

To explore this possibility, we briefly consider the standard paradigm for the
formation of the period gap.  As the secondary approaches the fully convective
boundary at the top of the period gap ($P_\mathrm{orb}\approx3$ hrs,
$M_\mathrm{2}\approx0.2-0.3$ \msun), magnetic braking is believed to be
disrupted as the tachocline, the interface region of strong shear between the
radiative and convective zones (and believed to be the site responsible for
generating the magnetic field) is lost, shutting down the dynamo
\citep[see][and references therein]{Browning}. Currently, however, there exists
evidence both for, and against, the disruption in magnetic breaking near the
convective boundary.  

\citet{PCEBs} found evidence for this disrupted magnetic breaking mechanism in
a survey of SDSS post common envelope binaries (PCEBs), examining the fraction
of PCEBs compared to non-interacting white dwarf plus main sequence binaries
(WDMS).  In the case of disrupted magnetic braking, there is a predicted steep
decrease in the number of PCEBs at the fully convective boundary due to their
much longer evolutionary timescales.  This was seen in their survey of SDSS
selected systems.  Also supporting the disrupted magnetic breaking scenario,
\citet{Browning} was only able to detect rotation ($v\sin{i}\geq2.5\
\mathrm{km\ s^{-1}}$) in seven out of 122 M dwarfs, of which four of these were
past the fully convective boundary. They suggest that this demonstrates that
magnetic breaking is less effective at the convective boundary. It is
interesting to note, however, that the seven stars with detected rotation all
showed high levels of stellar activity.

Conversely, \citet{AndronovPin} state that there is nothing ``magical'' about
the angular momentum loss rate or stellar activity levels at the fully
convective boundary.  This is supported by additional data for low mass stars
in clusters \citep[c.f.][and references therein]{Scholz}.  In fact,
\citet{Donati} show that there is a dramatic {\it increase} in the dynamo
generating processes at the fully convective boundary

Supposing that the secondary star becoming fully convective is not a reasonable
method for producing the period gap, is there an alternative method for
shutting down the magnetic braking? One possible method to offset the influence
of the high levels of magnetic activity seen at the fully convective boundary,
with the weaker magnetic braking observed for these stars, is to suppose that
the field topology changes from a predominantly large scale toroidal
configuration seen in stars with radiative zones \citep[see][]{Solanki}, to a
much more complex, non-axisymmetric topology in fully convective stars. In this
case, the global field might more closely resemble a collection of multi-polar
structures whose individual field strengths fall off rapidly with distance,
resulting in a much weaker global field than in the dipole case. Such fields
would have limited numbers of ``open field'' lines along which to transport
material, and thus much lower magnetic braking \citep[see][]{Cameron2002}. But
recent observations by \citet{Donati06} show that V374 Peg, a rapidly rotating
($P_\mathrm{rot}=10.69$ hr), fully convective M4V has a large scale
axisymmetric poloidal (dipole-like) field. A similar result was found for the
M4 dwarf GL 490B by \citet{PhanBao}.  Recent polarimetric observations across
the M dwarf sequence \citep{Donati, MorinMidMs, MorinLateMs} point towards a
sharp transition in magnetic field topology at 0.5 \msun, below which the
poloidal field topology dominates and its strength increases as mass decreases.
\citet{MorinLateMs} do note that it is possible to see wildly different
magnetic topologies between objects with similar stellar parameters, suggesting
that another parameter (perhaps age?) could play a role in the observed
topology.  In any case, there is strong evidence that the majority (85\%) of
the magnetic energy remains locked-away in smaller structures \citep{Reiners,
MorinMidMs}.  Perhaps it is the factor of $\sim 3$ increase (compared to V374
Peg) in rotational velocity for CV secondaries at the top of the period gap
that is sufficient to eliminate the large scale poloidal field and disrupt the
otherwise efficient magnetic breaking.  

One characteristic of rapid rotating late-type stars is that much of the
stellar activity is located closer to the poles of these objects
\citep[see][and references therein]{Bushby}. In some CVs polar spots are seen
\citep{Watson06}, while in others the spots are more equatorial
\citep{Watson07}. \citet{Cohen09} show that if a spot is located near the pole,
the stellar wind structure is dramatically affected, and that total mass and
angular momentum loss is substantially higher than if the spots were
equatorial. Unless it is found that the spots on CVs are preferentially located
near the equatorial regions, this explanation seems unable to produce the
desired result.

While observations of the magnetic field structures in low mass stars is
improving, as is our understanding of the generation of their magnetic dynamos,
the extension of this knowledge to explain the CV period gap is incomplete.
Since there is one, and only one property that all CVs at the top of the period
gap must share, a three hour rotation period, it must be that rotation somehow
quenches the efficient magnetic dynamo generation in fully convective stars.

\section{Conclusions}\label{conc}

We have performed a near-IR spectroscopic moderate resolution survey of nine
sub-gap CVs, detecting a signature of the secondary star in eight systems.  We
demonstrate an important link between abundance anomalies seen in the UV and in
the IR, as well as the reverse cases where the lack of abundance anomalies
appear.  Our clear detections show that future phase-resolved spectroscopy will
further constrain the nature of these secondary stars by obtaining radial
velocity measurements and therefore masses of the secondary stars in these
systems.  

While our results indicated a large fraction of sub-gap CVs contain a normal CO
abundance, we have insufficient objects to fully test this idea at present.
Ongoing analysis and modeling of these data presented here will allow us to
explore the CO bands seen in our spectra.  To fully assess the $^{12}{\rm
C}/^{13}{\rm C}$ ratio which gives the best indication and possibility to see
if the secondary star contains any CNO processed material \citep{COCNOref}, but
will require higher quality data.

We thank the anonymous referee for their useful and helpful comments.

This work is based on observations made with ESO Telescopes at the Cerro
Paranal Observatory under program ID 081.D-0225.

Some of the data presented herein were obtained at the W.M.  Keck Observatory,
which is operated as a scientific partnership among the California Institute of
Technology, the University of California and the National Aeronautics and Space
Administration. The Observatory was made possible by the generous financial
support of the W.M. Keck Foundation. The authors wish to recognize and
acknowledge the very significant cultural role and reverence that the summit of
Mauna Kea has always had within the indigenous Hawaiian community. We are most
fortunate to have the opportunity to conduct observations from this mountain.

This publication makes use of data products from the Two Micron All Sky Survey,
which is a joint project of the University of Massachusetts and the Infrared
Processing and Analysis Center/California Institute of Technology, funded by
the National Aeronautics and Space Administration and the National Science
Foundation.

RH would like to thank the New Mexico Space Grant consortium for Graduate
Research Fellowship support from 2007-09.

{\it Facilities:} \facility{VLT:Antu (ISAAC)}, \facility{Keck:II (NIRSPEC)}

\clearpage

\clearpage
\clearpage

\clearpage

\clearpage

\clearpage

\clearpage

\clearpage

\clearpage
\begin{deluxetable}{lcccccc}
\tabletypesize{\tiny}
\tablewidth{0.0pt}
\tablecaption{\sc{Observation Log}}
\tablecolumns{7}

\tablecaption{Observation Log \label{obslog}}

\tablehead{
        \colhead{System} &
        \colhead{Observation Date} &
        \colhead{Observation Time} &
        \colhead{Exp Time} &
        \colhead{\# Exp} &
        \colhead{Integration Time} &
        \colhead{Phase Covered} \\
   \colhead{\null} &
   \colhead{(UT)} &
   \colhead{(Midpoint, UT)} &
   \colhead{(s)} &
   \colhead{\null} &
   \colhead{(s)} &
   \colhead{(\%)}
}
\startdata
V2051 Oph     & 2008-05-12 & 07:51:30 & 70.0   & 30 & 2100 & 44.3   \\
V2051 Oph     & \nodata    & 08:54:18 & 70.0   & 30 & 2100 & 44.7   \\
V436 Cen      & 2008-05-13 & 00:54:22 & 70.0   & 30 & 2100 & 44.4   \\
V436 Cen      & \nodata    & 01:57:41 & 70.0   & 30 & 2100 & 44.9   \\
EX Hya (Keck) & 2005-02-17 & 12:53:57 & 240.0  & 4  & 960  & 13.4   \\
EX Hya        & 2008-05-12 & 06:03:27 & 25.0   & 36 & 900  & 21.4   \\
EX Hya        & 2008-06-21 & 02:00:42 & 25.0   & 36 & 900  & 20.9   \\
VW Hyi        & 2008-08-22 & 09:02:42 & 25.0   & 36 & 900  & 19.5   \\
VW Hyi        & \nodata    & 09:32:45 & 25.0   & 36 & 900  & 19.5   \\
Z Cha         & 2008-05-10 & 23:31:41 & 50.0   & 36 & 1800 & 33.3   \\
Z Cha         & 2008-09-17 & 09:12:36 & 50.0   & 36 & 1800 & 33.4   \\
WX Hyi        & 2008-06-21 & 08:47:54 & 40.0   & 36 & 1440 & 27.7   \\
WX Hyi        & \nodata    & 09:42:22 & 40.0   & 36 & 1440 & 27.5   \\
V893 Sco      & 2008-05-10 & 06:39:43 & 40.0   & 36 & 1440 & 27.3   \\
V893 Sco      & 2008-05-12 & 06:52:35 & 40.0   & 36 & 1440 & 27.2   \\
RZ Leo (Keck) & 2007-03-05 & 12:21:43 & 240.0  & 12 & 2880 & 44.6   \\
TY PsA        & 2008-06-20 & 06:59:29 & 70.0   & 30 & 2100 & 33.3   \\
TY PsA        & \nodata    & 08:04:50 & 70.0   & 30 & 2100 & 33.3   \\
\hline
LHS 427       & 2008-07-18 & 02:34:55 & 8.0   & 64 & 512  & \nodata \\
LHS 427       & \nodata    & 02:55:00 & 8.0   & 64 & 512  & \nodata \\
LHS 2347      & 2008-06-26 & 23:46:17 & 30.0  & 30 & 900  & \nodata \\
LHS 2347      & 2008-07-17 & 23:21:52 & 30.0  & 30 & 900  & \nodata \\
LHS 3003      & 2008-06-20 & 04:17:11 & 10.0  & 56 & 560  & \nodata \\
LHS 3003      & \nodata    & 03:35:16 & 10.0  & 56 & 560  & \nodata \\
\hline
\enddata
\tablecomments{All objects were observed at the VLT Antu telescope using ISAAC
with the exception of RZ Leo \& one observation of EX Hya as described in the
text.  ISAAC observations required two wavelength center positions to cover the
red half of the $K$-band, $2.25$ \& $2.35$ $\mu m$, whereas the Keck 
observations required only a single setting centered at $2.21\ \mu m$.}
\end{deluxetable}

\clearpage
\begin{deluxetable}{lccccccc}
\tabletypesize{\scriptsize}
\tablewidth{0.0pt}
\tablecolumns{8}
\tablecaption{Program Object Observed Properties \label{sysparams}}
\tablehead{
        \colhead{System} &
        \colhead{$P_\mathrm{orb}$\tablenotemark{a}} &
        \colhead{Primary Mass\tablenotemark{a}} &
        \colhead{Secondary Mass\tablenotemark{a}} &
        \colhead{Inclination\tablenotemark{a}} &
        \colhead{Magnitude} &
        \colhead{($J - H$)$_\mathrm{2MASS}$} &
        \colhead{($H - K$)$_\mathrm{2MASS}$} \\
   \colhead{\null} &   
   \colhead{(hrs)} &
   \colhead{(\msun)} &
   \colhead{(\msun)} &
   \colhead{(degrees)} &
   \colhead{($m_\mathrm{k}$)} &
   \colhead{\null} &
   \colhead{\null} \\
}
\startdata
V2051 Oph & 1.4982  & 0.78 $\pm$0.06  & 0.15$\pm$0.03   & 83  $\pm$2    & 13.530 & 0.46$\pm$0.05 & 0.34$\pm$0.06 \\
 V436 Cen & 1.5000  & 0.7  $\pm$0.1   & 0.17            & 65  $\pm$5    & 13.526 & 0.36$\pm$0.04 & 0.33$\pm$0.05 \\
   EX Hya & 1.6376  & 0.790$\pm$0.026 & 0.108$\pm$0.008 & 77.8$\pm$0.4  & 11.69  & 0.32$\pm$0.03 & 0.26$\pm$0.03 \\
   VW Hyi & 1.7825  & 0.67 $\pm$0.22  & 0.11 $\pm$0.03  & \nodata       & 11.702 & 0.49$\pm$0.03 & 0.34$\pm$0.03 \\
    Z Cha & 1.7880  & 0.84 $\pm$0.09  & 0.125$\pm$0.014 & 81.l$\pm$0.14 & 13.314 & 0.40$\pm$0.05 & 0.25$\pm$0.05 \\
   WX Hyi & 1.7955  & 0.9  $\pm$0.3   & 0.16 $\pm$0.05  & 40 $\pm$10    & 12.961 & 0.67$\pm$0.17 & 0.28$\pm$0.23 \\
 V893 Sco & 1.8231  & 0.89            & 0.17            & 72.5          & 12.68  & 0.24$\pm$0.03 & 0.28$\pm$0.04 \\
   RZ Leo & 1.8249  & \nodata         & \nodata         & \nodata       & 15.387 & 0.31$\pm$0.03 & 0.23$\pm$0.04 \\
   TY PsA & 2.02    & \nodata         & \nodata         & \nodata       & 13.583 & 0.42$\pm$0.05 & 0.29$\pm$0.05 \\
\hline
LHS 427   & \nodata & \nodata         & \nodata         & \nodata       & 6.734  & \nodata       & \nodata \\
LHS 2347  & \nodata & \nodata         & \nodata         & \nodata       & 12.038 & \nodata       & \nodata \\
LHS 3003  & \nodata & \nodata         & \nodata         & \nodata       & 8.928  & \nodata       & \nodata \\
\hline
\enddata
\tablenotetext{a}{CV data were taken from \citet[][and references therein]{RitterKolb}}.
\end{deluxetable}

\clearpage
\begin{deluxetable}{lcc}
\tabletypesize{\tiny}
\tablewidth{0.0pt}
\tablecolumns{3}
\tablecaption{Derived Spectral Types of Program Objects. \label{ourresults}}
\tablehead{
    \colhead{System} &
    \colhead{Derived Spectral Type} &
    \colhead{Published Spectral Type} \\
}
\startdata
V2051 Oph  & M7      & \nodata                \\
V436 Cen   & M8      & \nodata                \\
EX Hya     & M5      & M5-M6\tablenotemark{a} \\
VW Hyi     & M4      & L0\tablenotemark{b}    \\
Z Cha      & M5      & M5.5\tablenotemark{c}  \\
RZ Leo     & M4      & M5\tablenotemark{d}    \\
WX Hyi     & \nodata & \nodata                \\
V893 Sco   & $>$ M6  & \nodata                \\
TY PsA     & $>$ M6  & \nodata                \\
\hline
\enddata
\tablecomments{Estimated uncertianty is $\pm$ 1 spectral type for our determinations.}
\tablenotetext{a}{\citealt{BeuermannEX}}
\tablenotetext{b}{\citealt{Mennickent}}
\tablenotetext{c}{\citealt{WadeZCha}}
\tablenotetext{d}{\citealt{MennickentRZ}}
\end{deluxetable}

\begin{deluxetable}{lcccc}
\tabletypesize{\scriptsize}
\tablewidth{0.0pt}
\tablecolumns{5}
\tablecaption{CO Absorption Strength Across All CV Subtypes\label{biglist}}

\tablehead{
  \colhead{Star} &
  \colhead{Subtype} &
  \colhead{$P_\mathrm{orb}$ (hrs)} &
  \colhead{CO Ab. \tablenotemark{a}} &
  \colhead{Ref \tablenotemark{b}}
}
\startdata
\cutinhead{Pre-CV Systems}
P83l-57        & Pre-CV         & \nodata & Y             & 11  \\
HS1136         & Pre-CV         & 20.1    & ND            & 8   \\
RE 1016-053    & Pre-CV         & 18.9    & Y             & 11  \\
UZ Sex         & Pre-CV         & 14.3    & Y             & 11  \\
EC 12477-1738  & Pre-CV         & 13.7:   & Y             & 11  \\
V471 Tau       & Pre-CV         & 12.5    & Y             & 8   \\
EC 13349-3237  & Pre-CV         & 11.4:   & Y             & 11  \\
EC 14329-1625  & Pre-CV         & 8.4:    & Y             & 11  \\
BPM 6502       & Pre-CV         & 8.08    & Y             & 11  \\
RR Cae         & Pre-CV         & 7.29    & Y             & 11  \\
CC Cet         & Pre-CV         & 6.82    & Y             & 11  \\
SDSS0743       & Pre-CV         & 4.6     & Y             & 8   \\
BPM 71214      & Pre-CV         & 4.33    & Y             & 11  \\
BPM 71213      & Pre-CV         & 4.33    & Y             & 8   \\
EC 13471-1258  & Pre-CV         & 3.62    & Y             & 11  \\
LTT 560        & Pre-CV         & 3.54    & Y             & 11  \\
SDSS0757       & Pre-CV         & 3.5     & Y             & 8   \\
NN Ser         & Pre-CV         & 3.12    & Y             & 11  \\
SDSS0830       & Pre-CV         & 2.9     & Y             & 8   \\
\cutinhead{Non-Magnetic Systems}
EY Cyg      & DN UG          & 11.0 & W\tablenotemark{c}  & 9   \\
BT Mon      & NL SW          & 7.99 & ND                  & 8   \\
SY Cnc      & DN ZC          & 9.12 & ND\tablenotemark{*} & 5   \\
RU Peg      & DN UG          & 8.99 & W                   & 5   \\
CH UMa      & DN UG          & 8.23 & W                   & 5   \\
MU Cen      & DN UG          & 8.21 & W                   & 5   \\
AC Cnc      & NL SW          & 7.21 & Y ?                 & 5   \\
EM Cyg      & DN ZC          & 6.98 & W\tablenotemark{**} & 5   \\
V426 Oph    & DN ZC          & 6.85 & Y                   & 5   \\
SS Cyg      & DN UG          & 6.60 & W                   & 5   \\
AH Her      & DN ZC          & 6.20 & W                   & 5   \\
BV Pup      & DN UG          & 6.35 & ND                  & 5   \\
EX Dra      & DN UG          & 5.04 & Y                   & 4   \\ 
TW Vir      & DN UG          & 4.38 & N                   & 4   \\
SS Aur      & DN UG          & 4.38 & Y                   & 8   \\
U Gem       & DN UG          & 4.25 & W\tablenotemark{d}  & 4   \\
UU Aql      & NL SW          & 3.92 & N                   & 4   \\
IP Peg      & DN UG          & 3.80 & Y                   & 4   \\
RR Pic      & NL Nb SW       & 3.48 & W                   & 4   \\
TY PsA      & DN SU          & 2.02 & ND                  & 10  \\
RZ Leo      & DN SU          & 1.82 & Y                   & 8   \\
V893 Sco    & DN SU          & 1.82 & ?                   & 10  \\
WX Hyi      & DN SU          & 1.80 & ND                  & 10  \\
Z Cha       & DN SU          & 1.79 & W                   & 10  \\
VW Hyi      & DN SU          & 1.78 & Y                   & 10  \\
VY Aqr      & DN SU WZ       & 1.51 & N                   & 1   \\
V436 Cen    & DN SU          & 1.50 & Y                   & 10  \\
V2051 Oph   & DN SU          & 1.50 & Y                   & 10  \\
WZ Sge      & DN SU WZ       & 1.35 & E                   & 6   \\
GW Lib      & DN SU WZ ZZ    & 1.33 & ?                   & 8   \\
EI Psc      & DN SU          & 1.07 & N\tablenotemark{e}  & 1   \\
\cutinhead{Magnetic Systems}
GK Per      & DN Na IP       & 47.9 & W                   & 3   \\
AE Aqr      & NL DQ          & 9.86 & W\tablenotemark{f}  & 7   \\
V1309 Ori   & NL AM          & 7.98 & W\tablenotemark{g}  & 8   \\
MQ Dra      & NL AM LA       & 4.39 & Y                   & 3   \\
SDSS0837    & NL AM LA       & 3.18 & Y                   & 8   \\
AM Her      & NL AM          & 3.09 & Y                   & 4   \\
AR UMa      & NL AM          & 1.93 & Y                   & 3   \\
ST LMi      & NL AM          & 1.91 & Y                   & 3,8 \\
MR Ser      & NL AM          & 1.89 & Y                   & 4   \\ 
VV Pup      & NL AM          & 1.67 & Y                   & 2,3 \\
EX Hya      & NL IP          & 1.64 & Y                   & 10  \\
\hline
\enddata
\tablecomments{Only objects with NIR observations in the $K$-band with $R \gtrsim 1500$ are included.  A colon next to the 
orbital period indicates an uncertian result.}
\tablenotetext{a}{Y=appears normal for spectral type; W=appears weaker than
normal for spectral type; N=not present, but should have been for spectral type; ND=not detectable; ?=too low S/N;
E=emission.}
\tablenotetext{b}{(1) \citealt{TomVYEI}, (2) \citealt{SteveEF}, (3)
\citealt{TomPolars}, (4) \citealt{TomShorts}, 
(5) \citealt{TomLongPeriod}, (6) \citealt{SteveWZSge}, (7)
\citealt{TomPolarsSpitzer}, (8) \citealt{SteveRZLeo}, (9) \citealt{TomPrvComm},
(10) This Work (11) \citealt{clausprecv}}
\tablenotetext{c}{\citet{SionEYCyg, Boris2003}}
\tablenotetext{d}{\citet{UgemUV}}
\tablenotetext{e}{\citet{Boris2003}}
\tablenotetext{f}{\citet{AEAqrUV}}
\tablenotetext{g}{\citet{Szkody96, SchmidtStockman}}
\tablenotetext{*}{Very early spectral type, G1.5V so CO bands are not prominent}
\tablenotetext{**}{3rd light contanimation in the system, see \citet{North}.}
\end{deluxetable}

\end{document}